# Sequential Checking: Reallocation-Free Data-Distribution Algorithm for Scale-out Storage

Ken-ichiro Ishikawa, System Platform Research Laboratory, NEC

*Abstract*—Using tape or optical devices for scale-out storage is one option for storing a vast amount of data. However, it is impossible or almost impossible to rewrite data with such devices. Thus, scale-out storage using such devices cannot use standard data-distribution algorithms because they rewrite data for moving between servers constituting the scale-out storage when the server configuration is changed. Although using rewritable devices for scale-out storage, when server capacity is huge, rewriting data is very hard when server constitution is changed. In this paper, a data-distribution algorithm called Sequential Checking is proposed, which can be used for scale-out storage composed of devices that are hardly able to rewrite data. Sequential Checking 1) does not need to move data between servers when the server configuration is changed, 2) distribute data, the amount of which depends on the server's volume, 3) select a unique server when datum is written, and 4) select servers when datum is read (there are few such server(s) in most cases) and find out a unique server that stores the newest datum from them. These basic characteristics were confirmed through proofs and simulations. Data can be read by accessing 1.98 servers on average from a storage comprising 256 servers under a realistic condition. And it is confirmed by evaluations in real environment that access time is acceptable. Sequential Checking makes selecting scale-out storage using tape or optical devices or using huge capacity servers realistic.

## I. Introduction

SCALE-OUT storage is one option for large-volume storage because of its low cost, which increases linear to its capacity. Scale-out storage enables low-cost and large-volume storage by adding servers with low-cost devices.

Scale-out storage becomes more cost efficient and enables larger volume by using tape and optical devices, which are lower cost and have larger volume. Scale-out storage uses data-distribution algorithms for selecting a data-storing server for treating the vast amount of data in a typical architecture [1] [2]. Standard data distribution algorithms move data between servers when server configuration is changed. This causes rewriting data in scale-out storage. However, rewriting data in tape or optical devices is impossible in many cases or incurs high cost because it results in data fragmentation. Thus, scale-out storage with tapes or optical devices is not feasible with standard data-distribution algorithms. In a similar fashion, scale-out storage with servers having huge capacity is not practicable with standard data-distribution algorithms.

To solve this problem, a new data-distribution algorithm is necessary for such a scale-out storage. This algorithm is require to have the following characteristics for practical use: 1) the amount of data each server holds depends on its capacity, 2) the algorithm writes and reads data with realistic cost, and 3) it reads only the newest datum among ones having the same data ID.

In this paper, a data-distribution algorithm called Sequential Checking is proposed. It has the above characteristics and can be used for scale-out storage with devices which do not have rewrite function.

Sequential Checking has the following characteristics.
1. It does not move data when a server is added or amount of free capacity of a server is changed.
    Only adding a server and changing an un-used volume of a server are possible without rewriting data when the server configuration is changed. Because it is impossible for a server to be removed or delete the used volume of a server without data movement.
2. The amount of data written for each server is proportional to its amount of free space. This means that all servers become full when the storage becomes full.
3. It can select a unique server to write.
4. It can select a server(s) to read. They are few in most cases. And it can find out a unique server that stores the newest datum in them.

Therefore, Sequential Checking makes scale-out storage using tape or optical devices a reality.

The rest of this paper is structured as follows. Section 2 introduces Sequential Checking. Section 3 presents its characteristics, and Section 4 discusses its quantitative evaluation. Section 5 presents a discussion, and Section 6 shows related work. Section 7 provides a brief summary of this paper.

## II. Sequential-Checking Algorithm

This section discusses an algorithm of Sequential Checking. In Sequential Checking, there are four types of servers. Data-reading server is a server to be accessed to read datum.

Ken-ichiro Ishikawa, System Platform Research Laboratory, NEC, Kawasaki, Japan (e-mail: ishikawa@kikurage.net)

Data-writing server is a server to be accessed to write datum. Data-invalid server is a server to be accessed to invalid datum. Data-storing server is a server that stores datum. Sub-section 2A explains the basic concept of the algorithm, Sub-section 2B discusses parameter calculations, and Sub-section 2C presents a procedure of determining a data-writing server and data-invalid servers. Then Sub-section 2D introduces a procedure of determining a data-reading server(s), and Sub-section 2E introduces a data-duplication method.

*A. Basic concept of Sequential Checking*

This section explains the basic concept of Sequential Checking.

In typical architecture, scale-out storage uses data-distribution algorithms for determining the access server for scalability. The stored data may move to another server when the server configuration is changed because these algorithms may change an access server for writing or reading data when such change occurs. However, this causes a problem when scale-out storage is composed by servers with devices that is impossible or almost impossible to rewrite data.

To solve this problem, Sequential Checking does not move data between servers when the server configuration is changed. Instead, it searches each datum effectively when it reads data. It writes a datum to only limited servers, which depends on the data ID of the datum and server configuration. It reads only servers that may have written the datum. Thus, there are few reading servers in usual. Sequential Checking selects a server for writing datum depending on the un-used volume of that server. Thus, all servers become full when the storage becomes full in theory.

Figure 1 shows Sequential Checking, and the following sub-sections explain it in detail.

| Server Number | 0 | 1 | 2 | 3 | 4 | 5 | 6 |
|---|---|---|---|---|---|---|---|
| RAND | 0.81 | 0.73 | 0.24 | 0.37 | 0.55 | 0.12 | 0.18 |
| WriteP | 1.00 | 0.50 | 0.33 | 0.25 | 0.20 | 0.17 | 0.14 |
| Write or not | | | | | | | Write (Invalid) |
| Write first server if WriteP > RAND | | | | | | | |
| ReadP | 1.00 | 0.67 | 0.40 | 0.32 | 0.37 | 0.17 | 0.20 |
| Read or not | Read | | Read | | | Read | Read |
| Read every server if ReadP > RAND | | | | | | | |

Always ReadP ≥ WriteP

Fig.1 a brief algorithm of Sequential Checking

*B. Parameter setting*

Sequential Checking uses two parameters, WriteP for selecting a data-writing server, and ReadP for selecting a data-reading server(s). Sequential Checking changes both parameters when the storage configuration is changed. Sequential Checking assign number to servers. These numbers start from 0. The number assigned to a server is not changed. Server X is a server assigned X.

Both parameters are determined with the following procedure.

1. WriteP$_Y$ (WriteP for a server Y) is determined by the following equation where $V_X$ is the free space of server X.
    $$\text{WriteP}_Y = V_Y / \sum_{N=0}^{Y} V_N$$
2. ReadP$_Y$ (ReadP for server Y) is determined with the following algorithm. Its initial number is 0.0.
    If ReadP$_Y$ < WriteP$_Y$ then ReadP$_Y$ = WriteP$_Y$

Note that $0.0 \leq \text{WriteP}, \text{ReadP} \leq 1.0$, WriteP$_0$ = 1.0, and ReadP$_0$ = 1.0.

*C. Procedure for writing a datum*

Sequential Checking selects a data-writing server with the following procedure. When a datum is written, Sequential Checking invalids old data that have the same ID in some cases.

1. Each server has a pseudo-random number (RAND). A RAND's seed is calculated from the number assigned to a server and data ID. RAND satisfies $0.0 \leq \text{RAND} < 1.0$.
    For example, Mersenne Twister [3] [4] or XOR SHIFT [5] can be used for generating RANDs.
2. Sequential Checking compares the WriteP and RAND in descending order of server number. The first server that satisfies this equation is a data-writing server.
    WriteP$_{\text{server number}}$ > RAND
3. Sequential Checking compares the ReadP and RAND on servers assigned numbers that are greater than the number assigned a data-writing server. All servers that satisfy the following equation are data-invalid servers.
    ReadP$_{\text{server number}}$ > RAND

In Fig.1, for example, Sequential Checking compares the WriteP and RAND from server 6 to server 0. Since WriteP is larger than RAND in server 5, Sequential Checking selects the server as a data-writing server. It then compares the ReadP and RAND on all servers whose number is greater than 5 (there is only a server 6 in this case). Because ReadP is larger than RAND in a server 6, Sequential Checking then selects the server 6 as a data-invalid server.

The amount of data written to each server is proportional to the un-used volume of each server, as discussed in Sub-section 3A. Data invalidatoin is discussed in Sub-sections 5E and F.

*D. Procedure for Reading a datum*

Sequential Checking selects a data-reading server(s) with the following procedure.

1. Compute RAND. This step is the same as writing.
2. Sequential Checking compares ReadP and RAND in descending order of server number. Servers that satisfy the following equation are data-reading servers. This step proceeds until the datum is found.
    ReadP$_{\text{server number}}$ > RAND

In Fig.1, for example, Sequential Checking compares ReadP and RAND from server 6 to server 0. Since ReadP is larger than RAND in servers 6, 5, 2, and 0, Sequential Checking selects these servers as data-reading servers. Then as servers are read in descending order of the server number, the datum is found in server 5.

The proofs of being able to read the newest datum with this procedure are given in Sub-sections 3B and C.

*E. Data duplication*

Data are often duplicated for preventing data loss in scale-out storage. Sequential Checking duplicates data with the following idea.
1. When the duplication number is N, N servers form one group.
2. The algorithm works by treating one group as one server.
3. The datum is duplicated in the group.

When a datum is written, Sequential Checking selects a data-writing group and duplicates the datum in the group. Then it selects a data-invalid group(s) and invalids the datum in the group. When a datum is read, Sequential Checking selects a data-reading group(s), and the datum is read from a live server in this group.

III. PROOFS OF SEQUENTIAL-CHECKING CHARACTERISTICS

*A. Proof of proportionality between probability of writing datum on each server and volume of each server*

This section proves that the probability of writing a datum on each server is proportional to the volume of each server.

First, the proof focuses on a server assigned the largest number. The probability of selecting this server as the data-writing server can be calculated as

$V_Y / \sum_{N=0}^{Y} V_N$

V: Un-used volume of the server when parameters are set.
Y: a number assigned this server.

The ratio of un-used volume of this server to un-used volume of the storage is expressed as

$V_Y / \sum_{N=0}^{Y} V_N$.

Both expressions are the same, and Sequential Checking is guaranteed to always select a data-writing server because the WriteP of the server 0 is 1.0. This means that the probability of writing a datum on a server assigned the largest number is proportional to the ratio of the un-used volume of that server to the un-used volume of the storage when parameters are set.

When the proof focuses from a server assigned larger number to one assigned smaller number, it fits for all servers. Thus, the probability of writing a datum on each server is proportional to the un-used volume of each server when parameters are set.

This means that the amount of data written for each server is proportional to the volume of each server when storage becomes full.

The meaning of proportionality between the amount of data written for each server and volume of each server is discussed in Sub-section 5A.

*B. Proof that data-reading servers always include data-storing server*

This section proves that data-reading servers always include a data-storing server.

This means that a server that may be a data-writing server is always selected as a data-reading server.

The following inequality is always satisfied in all servers that may be data-writing servers in Sequential Checking.

$WriteP_{CUR} > RAND_{CUR}$

$WriteP_{CUR}$: WriteP of that server

$RAND_{CUR}$: RAND of that server

The following inequality is always established in all servers that may be selected for data-writing servers with Sequential Checking.

$ReadP_{CUR} (\geq WriteP_{CUR}) > RAND_{CUR}$

$ReadP_{CUR}$: ReadP of that server

This means that all servers selected for data-writing servers are chosen data-reading servers. Thus, data-reading servers always include a data-storing server if the storage stores that datum.

*C. Proof that Sequential Checking selects server that stores newest datum*

This section proves that Sequential Checking can select a server that stores the newest datum. Because the ReadP on each server is the maximum value of WriteP on each server, only servers that satisfy the following inequality can be data-writing servers.

$ReadP_{CUR} > RAND_{CUR}$

$ReadP_{CUR}$: ReadP of each server
$RAND_{CUR}$: RAND of each server

When a datum is written, servers assigned larger numbers than that of a data-writing server and satisfy the following inequality are selected as data-invalid servers.

$ReadP_{CUR} > RAND_{CUR}$

Sequential Checking then selects data-reading servers from servers assigned larger numbers to those assigned smaller numbers.

All servers that may be written an old datum and assigned larger numbers than that of a server that writes the newest datum become data-invalid servers. This means that a data-storing server has the newest datum when datum is read because Sequential Checking selects data-reading servers from servers assigned larger numbers to those assigned smaller numbers.

IV. EVALUATION OF SEQUENTIAL CHECKING

This section discusses the quantitative evaluation of Sequential Checking. Because a RAND generator is important for Sequential Checking, Mersenne Twister [4], which can generate almost homogeneous RANDs, was used for the evaluation.

*A. Confirmation and Quantitative evaluation of basic characteristics*

This sub-section discusses the confirmation of the following two basic characteristics of Sequential Checking through simulation.
1. The probability of writing a datum on each server is proportional to the volume of each server.
2. The newest datum can be read from a storage even when the un-used volumes of servers are changed.

First, I evaluated the probability of writing a datum on each server being proportional to the volume of each server. The case in which all servers have the same volume was first evaluated. The number of server sets was 6, their volumes were 100 TB, and their total volume was 600 TB. Size of datum was 1 GB. The IDs of all data were unique. Servers that had more data than their volume were ignored. The results are listed in Table I. The WriteP of each server is rounded off to two decimal places. The table shows that the data volume, which is the same as the amount of data written for each server, was almost proportional to the volume of each server.

TABLE I
VOLUME OF EACH SERVER AND AMOUNT OF DATA WRITTEN TO EACH SERVER WHEN VOLUME OF EACH SERVER IS SAME

| Number of Servers | WriteP | Data Volume (TB) |
|---|---|---|
| 0 | 1.000 | 99.683 |
| 1 | 0.500 | 100.295 |
| 2 | 0.333 | 100.119 |
| 3 | 0.250 | 99.596 |
| 4 | 0.200 | 100.142 |
| 5 | 0.167 | 100.165 |

The case in which the un-used volume of each server and number of servers are changed was evaluated. The un-used volume of each server was determined using a uniform random number (equal or greater than 0.0 and less than 1.0). The initial number of server sets was 2. One server was added after the un-used volume of each server was set and data were written twice. Maximum number of server is 6. The data IDs were unique, and the amount of data written to the storage was the un-used volume of the storage times 100,000. The determining of the un-used volume of servers and the writing of data to the storage was done a total of 10 times. Then the sum of un-used volume for each server (it was rounded off to two decimal places) and the amount of data written to each server were evaluated. The results are listed in Table II with final values of WriteP and ReadP (rounded off to one decimal place).

TABLE II
WRITEP, READP, VOLUME OF EACH SERVER, AND AMOUNT OF DATA WRITTEN TO EACH SERVER WHEN VOLUME OF EACH SERVER VARIED

| Number of servers | WriteP | ReadP | Total un-used volume of servers | Total amount of data written to servers |
|---|---|---|---|---|
| 0 | 1.00 | 1.00 | 5.695 | 567,632 |
| 1 | 0.41 | 0.77 | 5.611 | 561,938 |
| 2 | 0.16 | 0.73 | 3.298 | 331,479 |
| 3 | 0.03 | 0.29 | 1.660 | 165,652 |
| 4 | 0.19 | 0.40 | 2.031 | 202,758 |

Table II shows that the total amount of data written to each server is proportional to that of un-used volume for each server.

Finally, I examined that the newest datum could be read when the un-used volume of the server was changed. The number of servers increased from 1 to 6. The evaluation was repeated with an addition of one server and writing of data. The un-used volume of each server was set by a random number (equal or greater than 0.0 and less than 1.0). The amount of data was 1,000,000 per write sequence. Data IDs started as 0 and reset when the number of servers became 4. This means that data having the same data ID are written when there are 1 to 3 and 4 to 6 servers. After data writing finished, data were read, and it was confirmed whether the newest datum could be read.

TABLE III
WRITEP (FIRST WRITE), WRITEP (SECOND WRITE), READP AND WRITEP (READ), AND RAND (WHOLE CASES) FOR DATA WHOSE DATA ID WAS 1,234,567

| Number of Server | First Write WriteP | Second Write WriteP | Read ReadP | Read WriteP | RAND |
|---|---|---|---|---|---|
| 0 | <u>1.00</u> | 1.00 | <u>1.00</u> | 1.00 | 0.49 |
| 1 | 0.73 | 0.46 | 0.77 | 0.77 | 0.89 |
| 2 | | 0.15 | 0.68 | 0.68 | 0.02 |
| 3 | | 0.38 | 0.38 | 0.17 | 1.00 |
| 4 | | <u>0.22</u> | <u>0.22</u> | 0.06 | 0.09 |
| 5 | | | 0.20 | 0.20 | 0.59 |

Example results are listed in Table III, which shows WriteP (first write and second write), ReadP and WriteP (read), and RAND (whole cases) when the data ID of the accessing datum was 1,234,567. The underlined numbers are the WriteP of data-writing servers or ReadP of data-reading servers. The RAND of a server having a server number of 3 became 1.00 due to a rounding error. From this evaluation, it was confirmed that only the newest data can be read.

*B. Quantitative evaluation of Sequential Checking*

This section discusses the quantitative evaluations of Sequential Checking through simulation. The first evaluation involved measuring the error of proportion between the amount of data written to each server and un-used volume of each server. The second evaluation involved the number of data-reading servers when a datum was read in an assumed case of actual usage.

*1) Error of proportion between amount of data written to each server and un-used volume of each server*

An error of proportion between the amount of data written to each server and the un-used volume of each server was evaluated. The number of servers was 256, and the un-used volume of each server was set by a uniform random number (equal or greater than 0.5 and less than 1.5). The number of data is size of un-used volume of storage times 1,000,000. This evaluation involved measuring the maximum error between the amount of data actually written and that estimated from the un-used volume of each server.

The results show that the maximum error was from 0.28 to 0.55%, and average maximum error was 0.35%. The low maximum error makes full use of the availabe storage volume. If the maximum error is high, the storage becomes unwritable, even if there are large un-used volumes in other servers, because the fullness of one server makes the entire storage un-writable.

*2) Number of data-reading servers when datum is read*

The second evaluation involved measuring the number of data-reading servers when a datum was read. The following were estimated in the case of actual usage.

1. Increase in volume of at least server that was added when a certain percentage of the storage became full
2. Addition of a server when the volume of a server that was previously added becomes maximum
3. Write data until the storage becomes full after the last server is added and the volume of all servers becomes maximum

After all data are written, the number of data-reading servers and data-reading servers that were accessed until the datum was found were measured.

The initial number of a server was 1, the initial volume of a server was 100 TB, and the maximum volume of a server was 1 PB. When the volume of a server increased, 100 TB was added, and when the number of servers increased, one server was added. The increase in volume or increase in servers was done when the storage volume was filled by $N \times 100\%$ (N is from 0.0 to 1.0 in 0.1 increments). The maximum number of servers was from 1 to 256. Then the size of a datum was 1 GB. Because all data written after the storage expansion ends can be read by one access, the following three cases were evaluated.

1. Number of data-reading servers

2. Number of servers accessed when data are read (only data written before the storage expansion ends are evaluated)
3. Number of servers accessed when data are read (all data are evaluated)

*a) Number of data-reading servers*

Figure 2 shows the number of data-reading servers. The results are out of the figure when N is 0.8 or greater to show the difference of the number of data-reading servers when N is small. For example, when N was 0.5, which means the storage volume increased when half of it became full, the number of data-reading servers was less than 11 when the maximum number of servers was 256.

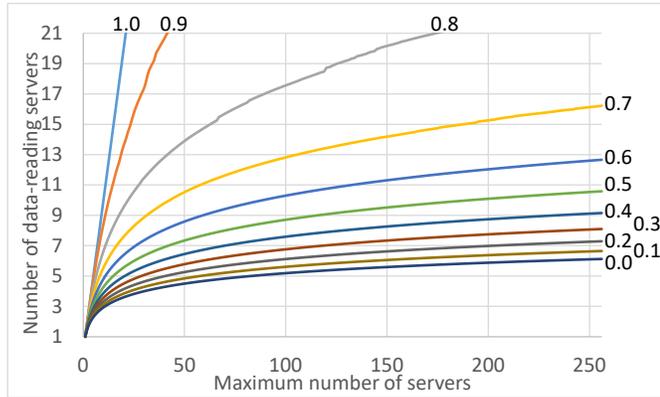

Fig. 2. Number of data-reading servers

*b) Number of servers accessed when data are read (evaluated data are those written only before storage expansion ends)*

Figure 3 shows the number of servers accessed when data were read. The data used in this evaluation were limited to those only written before storage expansion ended. The results are out of the figure when N is 0.8 or greater to show the difference of the number of servers accessed when N is small. For example, when N was 0.5, the number of servers accessed was less than 3 when the maximum number of servers was 256. When the maximum number of servers was over 30, the number of servers accessed hardly increased by increasing the maximum number of servers. This shows the significant scalability of Sequential Checking.

*c) Number of servers accessed when data are read (all data are evaluated)*

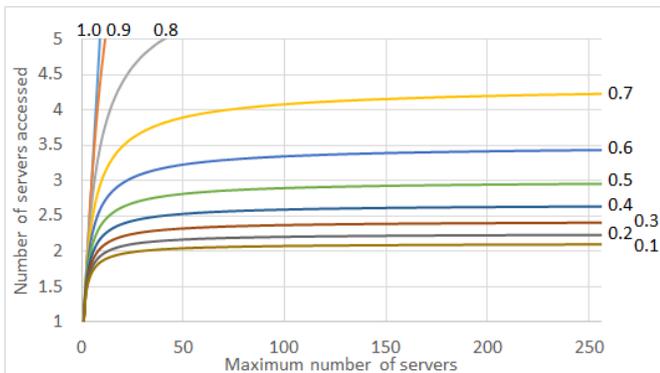

Fig. 3. Number of servers accessed when data were read
(Evaluated data were those written only before storage expansion ended)

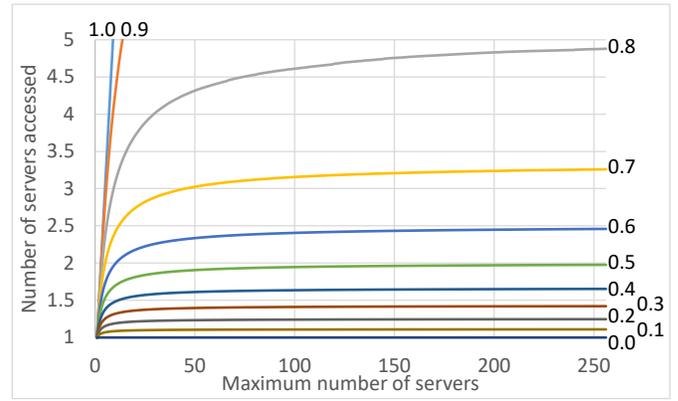

Fig. 4. Number of servers accessed when data were read
(All data were evaluated)

Figure 4 shows the number of servers accessed when data were read. All data were used in this evaluation. The results are out of the figure when N is 0.9 or greater to show the difference of the number of servers accessed when N is small. For example, when N was 0.5, the number of servers accessed was only around 2 when the maximum number of servers was 256.

From Figs. 2, 3, and 4, lower threshold of data volume in a storage for increasing volume of the storage makes number of data-reading server and servers accessed smaller. However, if threshold of data volume in a storage for increasing volume of the storage is lower, the storage cost become higher. Thus, the threshold must be set based on performance and cost settings.

C. *Evaluation in real environment*

The access time of a storage using Sequential Checking and that of a storage using a standard data-distribution algorithm were measured to evaluate the effectiveness of using Sequential Checking. Consistent Hashing [25] was used as the standard algorithm. A client server writes and reads data from/to Memcached in 8 storage servers, which were connected to the same hub.

The evaluation environment was as follows.
Server:
  CPU: INTEL Xeon X5550 2.67 GHz
  Memory: 24 GB, Network: 1000BASE-T
  Client server: 1 server, Storage server: 8 servers
Data:
  Size: 1 KB, Number of data: 8,000,000
  ID: integer from 0
Client software:
  OS: CentOS 6.3, Library: libmemcached 1.0.18
  Data-distribution algorithm: Consistent Hashing
                (Virtual Server 100)
              Sequential Checking
Storage software:
  OS: CentOS 6.3
  Software: Memcached 1.4.4 (Option -d -m 4096)

The reason of using Memcached in storage servers is as follows. When a datum is accessed from a server, the address of the datum is read from Device-A, then the datum is accessed from Device-B. When the datum is accessed from the storage, the datum is accessed only one time from/to Device-B. The time of accessing the datum from Device-B largely depends on the condition of Device-B because this paper assumes it is a tape device and so on. Thus, in this evaluation, the access time of Device-B was eliminated. Memcached was used as Device-A because a memory device was assumed as this device. The difference between access time in a real environment and that in this evaluation environment was the transfer-data time through a network and access-data time from Device-B. These do not differ by data-distribution algorithm.

The following cases were evaluated.
1. Write all data to a storage using Consistent Hashing
2. Read all data from a storage using Consistent Hashing
3. Write all data to a storage using Sequential Checking
4. Read all data from the following storage using Sequential Checking
    a. Repeat writing data 500,000 times and adding 1 server until the number of servers becomes 8
       The storage starts 1 server
    b. Write data 500,000 + 4,000,000 times to the storage
    c. Read all data from the storage

Evaluations 1 and 2 were conducted to estimate normal storage using Consistent Hashing. Evaluation 3 was conducted to estimate normal storage using Sequential Checking. Evaluation 4 was conducted to also estimate the normal use case of Sequential Checking, in which a server is added when the un-used volume of the storage becomes half. The amount of accessed data was the same in all evaluations. The evaluation results are the average of each evaluation conducted five times

TABLE IV
EVALUATIONS IN REAL ENVIRONMENT

| Test | Time (s) |
|---|---|
| Consistent Hashing (Write) | 1345.3 |
| Consistent Hashing (Read) | 1253.4 |
| Sequential Checking (Write) | 1321.8 |
| Sequential Checking (Read) | 1649.5 |

and listed in Table IV.

When data were written, the difference between the number of accesses to the server when the storage used Consistent Hashing and that to the server when the storage used Sequential Checking was only the number of invalidate commands by the storage using Sequential Checking. The invalidate command was not issued in this evaluation, which means that the difference depends on the execution time of the data-distribution algorithms. In this evaluation, the difference between execution time in the storage using Consistent Hashing and that in the storage using Sequential Checking was scope of measurement error.

When data were read, the number of read commands to the storage using Consistent Hashing was 8,000,000 and that to the storage using Sequential Checking was around 11,400,000. The execution time of the storage using Consistent Hashing was around 0.7 times that of the storage using Sequential Checking.

The maximum variability in the amount of data between servers was around 23.3% in the storage using Consistent Hashing and around 0.2% in the storage using the Sequential Checking. This means that the storage using Sequential Checking could use server volume far more efficiently than that using Consistent Hashing.

When the number of servers in the storage was small in a real environment, the execution time of Sequential Checking was mostly the same as that of Consistent Hashing. The overhead of Sequential Checking was only the increase in commands.

V. DISCUSSION

*A. Amount of data written to each server is proportional to volume of each server*

The server volume can be used fully when the amount of data written to each server is proportional to the volume of each server.

Sequential Checking is used for storages using devices that reusing un-used volume created by deleting data is difficult. Thus, used volume of a server depends on only the size of all data written to this server.

Therefore, if there are enough data, the average size of data becomes the same in the law of large numbers, distribution of the amount of written data becomes uniform in the law of large numbers, and the amount of data written to each server is proportional to the data volume stored in each server. Thus, server volume can be used fully because the amount of data written to each server is proportional to the volume of each server. This means that all servers become full when the storage becomes full when using Sequential Checking. The law of large numbers is frequently used in other data-distribution algorithms [25] [26] [27] [28] [29].

*B. Difference in frequency of reading between servers*

In a storage using Sequential Checking, at first, the frequency of reading data from servers that are added late becomes high compared with that estimated from the amount of data that the servers store, and secondly a server 0 is always selected as a data-reading server.

The first problem is not serious. For example, if a server is added for a storage when half of the storage becomes full, the frequency of reading data from servers added late is only double that of reading data stored in these servers when they are full.

The second problem is also not serious. A case of actually reading data from a server 0 is only that of reading data stored in this server or reading data not stored in this storage. Thus, the problem of concentrated accesses does not occur.

*C. Unstable latency of reading due to unstable number of accessed servers*

In a storage using Sequential Checking, the number of accessed servers is small in most cases when a datum is read. However, it is possible to access many servers in rare cases, which dramatically increases access time. This phenomenon cannot be prevented because of the characteristics of Sequential Checking. However, the storage devices using Sequential Checking are assume to be tape or optical. Such device stores cold data, which do not require short latency. This means that this problem is not serious.

If storage using Sequential Checking uses devices composed of RAM, SSDs, and HDDs, which store data that require short latency, system architect minds that latency can change dynamically.

## D. Timing of writing datum and number of access servers when this datum is read

In a storage using Sequential Checking, a datum written more recently is more frequently stored in a server assigned a large server number. This means that the number of accessed data-reading servers is smaller when the accessed datum is stored to the storage later than the number of accessed data-reading servers when the accessed datum is stored to the storage early. A datum written to the storage after the parameter is set last and can be read by the first access. This characteristic of Sequential Checking fits characteristics of data that data written recently are read more frequently.

## E. Frequency of invaliding datum when it is written

In storage using Sequential Checking, there is a case in which invaliding a datum is necessary when it is written. However, this case is rare.

In storage using Sequential Checking, an invalidate command is issued in the case of ReadP > WriteP (not ReadP = WriteP). This occurs when WriteP is decreased, and the decrease in WriteP results in the following cases.
1. The volume of that server is decreased.
2. The volume is added to a server assigned a smaller server number than the server number of that server.
3. The average size of data written to that server is larger than the average size of data written to servers assigned smaller server numbers.

In normal usage, server volume increases monotonically. A server is added when the volume of all servers becomes maximum. Thus, cases 1 and 2 never occur in normal usage. Case 3 may occur. However, the amount of decrease in WriteP is very small in most cases due to the distribution of the size of data written for each server becoming the same in accordance with the law of large numbers. Thus, the case of invaliding a datum when it is written is rare.

## F. Datum invalidation when it is written

As mentioned above, Sequential Checking is a data-distribution algorithm for storage using devices that cannot or are hardly able to delete data. Thus, data are not moved between servers when the server configuration is changed. However, a datum may be invalided when it is written in the storage using this algorithm.

This is not a servious problem. In a storage using a standard data-distribution algorithm, the volume of data moved between servers when the storage configuration is changed is 1/N that of the server when number of servers in the storage becomes N and the volume of the servers is the same. It is not acceptable that such a large server volume becomes unusable. However, invalidating data, when the datum is written in a storage using Sequential Checking, is rare from the discussion in Sub-section 5E. This means that volume waste in a storage is very small and does not cause a problem.

The method of invalidating data from un-rewritable devices depends on the system and device.

## VI. RELATED WORK

There are two major systems that distribute data to storages by using data distribution algorithms. These are RAID and distributed storage. Thus, RAID and other data-distribution algorithms for distributed storage are discussed. Storage for large data volumes were introduced in this paper because the main focus of Sequential Checking is such storage.

RAID [6] is data management technique for quicker access or more reliable store. It is a major research topic today [7] [8] [9]. Basically RAID is a technique for magnetic disk drives [10] [11] [12], but RAID for SSD and so on is also hot research area recently [13] [14] [15]. Erasure coding is similar technique for more reliability with RAID. Erasure coding can restore data from parts of the data modified. Research of coding technique has long history since 1970s [16] [17] [18] [19] to now [20] [21] [22].

In this paper, data-distribution algorithms with which only a minimum amount of data move between servers when the storage configuration is changed were considered because, to the best of the author's knowledge, standard data distribution algorithms assume that the storage can be rewritten data easily.

RUSHp [23] [24] is a similar data-distribution algorithm to Sequential Checking. RUSHp determines parameters and selects an access server in a similar way to Sequential Checking. However, RUSHp requires minimum data movement when a server is added. Thus, there are many differences in algorithms and characteristics between Sequential Checking and RUSHp.

Consistent Hashing [25] is the de-facto standard data-distribution algorithm for scale-out storage. It can distribute data uniformly among servers and requires a minimum amount of data movement between servers when a server is added or removed. Hash numbers calculated from servers are set on the hash ring. The area from the position of hash number of a server to that of another server in a certain direction on the hash ring is the area of that server. When a datum is accessed, the hash number of the datum is set on the hash ring, and the owner of the position of the datum on the hash ring is a data-storing server. Random Slicing [26] is a version of Consistent Hashing. With this algorithm, the hash line is divided depending on the server volume. It can distribute data to servers more uniformly than Consistent Hashing.

Highest Random Weight [27] can distribute data uniformly to servers and requires minimum data movement when a server is added or removed. With this algorithm, a datum is written to a server whose hash number that is calculated from the data ID and the server ID is maximum between the servers. The subspecies of Highest Random Weight are Straw Buckets in CRUSH [28] and Weighted Rendezvous Hashing [29]. They can be used for storages composed of servers having un-uniform capacity.

Freeze-ray [30] is a representative storage composed of optical devices. Lustre [31], Ceph [32], Google File System [33], and Hadoop Distributed File System [34] are representative massive storage systems and Amazon S3 [35], Google Cloud Storage [36], and Microsoft Azure Storage [37] are representative cloud storage servers. Then Amazon Glacier [38] and Google Cloud Storage Nearline [39] are representative cold storage services that have huge capacities.

## VII. CONCLUSION

In this paper, Sequential Checking was proposed, which is a data-distribution algorithm for scale-out storage composed of devices in which data movement between servers when the server configuration is changed is impossible or almost impossible. Sequential Checking has the following characteristics.

1. Data do not move between servers when the server configuration is changed.
2. The amount of data written to each server is proportional to the volume of each server. This means that all servers become full when the storage becomes full.
3. Only one data-writing server is selected when the datum is written, and a data-invalid server(s) is selected when the datum is written in rare cases.
4. The newest datum can be read by reading only few data-reading servers when a datum is read in most cases.

Through proofs and simulations, the basic characteristics of Sequential Checking were proved, and its performance was evaluated. As a result, 1.98 servers on average can be accessed when a datum is read in the setup condition, which is evaluated in this paper when the storage has 256 servers.

Sequential Checking enables scale-out storage composed of tape or optical devices or huge capacity servers which was previously almost impossible.

## APPENDIX A SEQUENTIAL CHECKING WITHOUT DATUM INVALIDATION

Using Sequential Checking without datum invalidation is possible by limiting the change of server configuration.

In this Sequential Checking (Sequential Checking Limited), available change of server configuration limits only the addition of a server and increase in the volume of the server added to the storage last. This limitation is not a problem in the normal use case of Sequential Checking.

The characteristic of Sequential Checking Limited compared with Sequential Checking are as follows.
1. Only an addition of a server or an increase in the volume of a server added last is available.
2. WriteP and ReadP becomes the same single parameter.
3. Only the parameter of a server added last is changed when server configuration is changed.

It is obvious that a parameter for only a server added last must be changed when the server configuration is changed because of the limitation of change of server configuration, an algorithm for calculating WriteP and ReadP in Sequential Checking, and law of large numbers. The WriteP and ReadP in Sequential Checking can be combined into one parameter due to WriteP monotonically increasing because the storage configuration is limited to an addition of a server or an increase in the volume of a server that is added last. Because ReadP=WriteP, invaliding a datum is not necessary when it is written. It is clear that the newest data can be read from a storage using Sequential Checking Limited by comparing algorithm of Sequential Checking.

Both Sequential Checking and Sequential Checking Limited have advantages and disadvantages. Thus, a system architect can select one of them depending on the system requirement.

## APPENDIX B DELETION OF SERVERS AND DEVICES STORING DATA

A storage using Sequential Checking can delete servers or devices storing data by use of a special method.

When a server is deleted from a storage using Sequential Checking, the WriteP and ReadP of that server are set to 0.0 and those of other servers are recalculated. The data stored in the deleted server are then re-distributed to another server.

A device storing data can be deleted in the same way as deleting server.
However, these methods weaken the characteristics of Sequential Checking. Thus, they are out of the scope of this paper.